\documentclass[onecolumn,aps,floatfix,pra,superscriptaddress,nofootinbib]{revtex4}

\usepackage{color}
\usepackage{graphicx}
\usepackage{bm}
\usepackage{amsmath}
\usepackage{amsbsy}
\usepackage{hyperref}
\usepackage{enumerate}
\usepackage{upgreek}
\usepackage[subnum]{cases}
\usepackage{dsfont}

\newcommand{\cl}{ \text{cl} }

\newcommand{\tr}{ \text{tr} }
\newcommand{\ti}{ \tilde }
\newcommand{\ep}{ \epsilon }
\newcommand{\pa}{ \partial }
\newcommand{\hb}{ \hbar }
\newcommand{\si}{ \sigma }
\newcommand{\om}{ \omega }

\newcommand{\la}{ \langle }
\newcommand{\ra}{ \rangle }
\newcommand{\del}{ \delta }
\newcommand{\Del}{ \Delta }

\newcommand{\be}{ \beta }
\newcommand{\lam}{ \lambda }

\newcommand{\erf}{ \text{erf} }
\newcommand{\erfi}{ \text{erfi} }
\newcommand{\nc}{ \text{nc} }
\newcommand{\thb}{ \tilde{\hbar} }
\newcommand{\tpsi}{ \tilde{\psi} }
\newcommand{\tPsi}{ \tilde{\Psi} }

\newcommand{\tvrho}{ \tilde{\varrho} }
\newcommand{\tu}{ \tilde{u} }
\newcommand{\tE}{ \tilde{E} }
\newcommand{\tC}{ \tilde{C} }

\newcommand{\tbe}{ \tilde{\beta} }
\newcommand{\trho}{ \tilde{\rho} }
\newcommand{\tG}{ \tilde{G} }
\newcommand{\tH}{ \tilde{\hat{H}} }
\newcommand{\tU}{ \tilde{\hat{U}} }
\newcommand{\tc}{ \tilde{c} }
\newcommand{\tA}{ \tilde{A} }
\newcommand{\tS}{ \tilde{S} }
\newcommand{\tsi}{ \tilde{\sigma} }

\newcommand{\Ai}{ \text{Ai} }
\newcommand{\wti}{ \widetilde }
\newcommand{\QM}{ \text{q} }
\newcommand{\fr}{ \text{f} }

\newcommand{\bsi}{ \bar{\sigma} }

\newcommand{\im}{ \text{Im} }

\begin{document}

\title{Scaled quantum theory. The bouncing ball problem }

%{ Bouncing ball in a smooth quantum-classical transition}

\author{S. V. Mousavi}
\email{vmousavi@qom.ac.ir}
\affiliation{Department of Physics, University of Qom, Ghadir Blvd., Qom 371614-6611, Iran}
\author{S. Miret-Art\'es}
\email{s.miret@iff.csic.es}
\affiliation{Instituto de F\'isica Fundamental, Consejo Superior de Investigaciones Cient\'ificas, Serrano 123, 28006 Madrid, Spain}
\begin{abstract}

Within the so-called scaled quantum theory, the standard bouncing ball problem is analyzed under the
presence of a gravitational field and harmonic potential. 
In this framework, the quantum-classical transition of the density matrix is described by the linear scaled von Neumann equation for mixed states and after it has been particularized 
to the case of pure states. The main purpose of this work is to show how this theory works for 
conservative systems and  the quantum-classical transition is carried out in a continuous and smooth
way, being equivalent to a nonlinear differential wave equation which contains a transition parameter ranging continuously from one to zero and covering all dynamical regimes in-between the two extreme quantum and classical regimes. This parameter can be seen as a degree of quantumness where all intermediate dynamical regimes show quantum features but are fading gradually when approaching to the classical value.

\end{abstract}

\maketitle


%%%%%%%%%%%%%%%%%%%%%%%%%%%%%%%%%%%%%

%========================================
%========================================

\section{Introduction}

In the sixties of the last century, Schiller \cite{Schiller1, Schiller2, Schiller3} and Rosen \cite{Rosen1, Rosen2} proposed and discussed what is known as the classical wave function ruled by a nonlinear differential wave equation. The nonlinearity was precisely due to the so-called quantum potential of Bohmian mechanics \cite{Holland-book-1993,Salva1,Salva2}. 
As mentioned by Holland \cite{Holland-book-1993}, 
a nonlinear differential equation may be a limiting case of a linear equation, the time-dependent 
Schr\"odinger equation. As a step forward to this proposal, Richardson et al.
\cite{RiSchMaVaBa-PRA-2014} introduced a nonlinear quantum-to-classical transition  differential wave equation by means of a transition parameter $\epsilon$ with $\epsilon \in [0,1]$, seen as a degree of quantumness, where the 
classical regime ($\epsilon =0$), in the sense of Schiller, was the classical differential wave equation and the quantum
regime ($\epsilon =1$) was the standard Schr\"odinger wave equation. Thus, this transition differential wave equation 
was a way to go from the quantum to classical regime in a smooth and continuous way. In other words, 
one is also able to describe any in-between dynamical regime  for a 
given quantum problem.  They also proved the equivalence of this nonlinear differential equation with a linear one, 
what we have termed the scaled Schr\"odinger equation, being the central differential wave equation of 
what we could call the {\it scaled quantum theory} or {\it scaled quantum mechanics}. Typical quantum features
are observed for any value of $\epsilon$ except for $\epsilon = 0$: discrete eigenvalues, interference, 
tunneling, and so on. The interesting point is that quantum features are observed in all dynamical regimes but they are fading gradually and continuously when decreasing
this parameter up to the strict classical value.
This new approach reminds us  the well-known WKB approximation which is based on a series expansion in powers of Planck's constant, widely used for conservative systems. Several important differences are worth mentioning.
Firstly, whereas the classical Hamilton-Jacobi equation for the  action is obtained 
at zero order in the WKB approximation,  the so-called classical wave (nonlinear) differential
equation \cite{Schiller1,Schiller2,Schiller3} is reached by construction. 
Furthermore, the hierarchy of the differential equations for the classical action at 
different orders of the expansion in $\hbar$ is substituted, in our case, by only a
transition wave differential equation which can be  solved in the linear domain following 
the standard numerical methods. Second,  the transition from quantum to classical
trajectories is carried out in a continuous and gradual way, much easier than the 
one followed by the WKB approximation. 
Third, the scaling procedure extended and applied to open quantum systems is very easy 
to implement. And fourth, the gradual decoherence process 
due to the variation of the scaled Planck's constant allows us to analyze what happens 
at intermediate dynamical regimes;  the transition parameter pinpointing and 
stressing different dynamical regimes until reaching gradually and continuously 
the classical regime. Finally, in the literature, several attempts apart from the WKB approximation have also 
been proposed for conservative systems based on the Wigner distribution function \cite{Wigner}, on large quantum numbers \cite{Kiwi} and on quantum 
tomograms \cite{Manko}. Our approach can be seen as an alternative and easy way 
to implement the quantum-classical transition because the scaled equations of motion, 
that is, the corresponding Schr\"odinger and the Liouville-von Neumann equations 
have the same structure than the standard ones and are 
solved using the same numerical techniques.

This new approach can be 
applied to any quantum problem as the standard quantum mechanics does. Interesting and simple applications 
of this transition differential wave equation are found with Gaussian wave packets \cite{Chou1,Chou2}. 
More recently, stochastic Bohmian and scaled trajectories have also been discussed in the literature 
for open quantum systems \cite{MoMi-AP-2018,MoMi-JPC-2018,MoMi-FOP-2022}. Bohmian and 
scaled trajectories are expressed as a sum of  a classical trajectory (a particle property) and a term 
containing the width of the corresponding wave packet (a wave property) within of what has been 
called the {\it dressing scheme} \cite{NaMi-book-2017}. As a new extension of this approach, in trying to describe classical mechanics with the language of quantum mechanics, we have recently established the scaled Liouville-von Neumann equation for the density matrix \cite{MoMi-Sym-2023}. Thus, diagonal elements of the density matrix, even the classical one, have probabilistic interoperation while the non-diagonal elements imply coherences.
This extension could open up the possibility to apply this scaled theory to nonequilibrium statistical mechanics.

In this work, our main goal is to apply the scaled quantum theory to a conservative system.
The so-called  bouncing ball (BB) problem has been extensively studied in quantum mechanics from both 
theoretical and experimental point of view. Different aspects of the quantum bouncer have been 
used; some of them have been developed for pedagogical purposes \cite{Gi-AJP-1975}. The dynamics of a quantum 
BB described initially by a Gaussian wave packet well-localized in a point far from the hard wall has 
been considered by examining collapses and revivals of oscillations \cite{Ba-AJP-1999}; and by analyzing 
the expectation values and uncertainties of position and momentum variables \cite{DoRo-AJP-2001}. 
See also \cite{Va-AJP-2000} and \cite{Go-AJP-1999} for some analytical results. Revivals of 
quantum wavepackets have been reviewed in \cite{Ro-PR-2004}. The role of the inertial and gravitational 
masses of a BB has been discussed in terms of both classical and quantum dynamics \cite{FeBeFe-EJP-2019}.
Some researchers have considered bouncing from a flat surface in the absence of gravity 
\cite{DoRo-EJP-1999, BeDoRo-PS-2005} or in the presence of a harmonic oscillator 
potential \cite{LeNg-EJP-2009} instead of the gravity. 
Gaussian-Klauder coherent states have also been examined for the quantum bouncer and the 
quantum-classical correspondence have been established for such states \cite{MaFo-PRA-2006}.   
The quantum gravitational BB has been carried out with atomic clouds \cite{Boetal-PRL-1999} and 
neutrons \cite{Neetal-Nature-2002, Neetal-EPJC-2005, JeStAbGe-NIMPRA-2009, 
AbJeStGe-NPA-2009, Abetal-PP-2011}. An optical analog of a quantum BB has been 
experimentally demonstrated by using a circularly curved optical waveguide \cite{Deetal-PRL-2009}.

This work is organized as follows. Section II introduces the quantum-to-classical smooth transition in terms of the density matrix which is ruled by the Liouville-von Neumann equation and Setion III applies this theoretical formalism to simple interaction potentials. 
%Section IV introduces the scaled time-independent 
%Schr\"odinger equation and the scaled harmonic oscillator is studied in this context by analyzing 
%its eigenvalues and eigenfunctions. 
Section IV deals with the scaled BB, described initially by a well-localized Gaussian wave packet, both under the presence of a gravitational field and also for a harmonic potential. 
%In the first case, BB is initially described by a well-localized Gaussian wave packet while, in the second case, the BB is also analyzed when it is initially described by a coherent state. 
In Section V, Ehrenfest relations are presented in this context and  
the concept of non-classical force reminiscent of the quantum effective force is defined, which has already 
been introduced in the realm of quantum mechanics. Finally, in
the last section, some conclusions about the scaled quantum theory applied to conservative problems are
summarized.

\section{Quantum-classical smooth transition. The scaled von Neumann equation}

Following \cite{RiSchMaVaBa-PRA-2014} we have proposed \cite{MoMi-Sym-2023} to describe both quantum and classical mechanics with the same language i.e., the density matrix. Using the polar form of the density matrix in the von Neumann equation led us to propose the {\it nonlinear} classical von Neumann equation
\begin{eqnarray} \label{eq: cl-vN}
i\hb \frac{\pa}{\pa t} \rho_{\cl}(x, y, t) &=& - \frac{\hb^2}{2m} 
\left( \frac{\pa^2}{ \pa x^2 } - \frac{\pa^2}{ \pa y^2 }   \right) \rho_{\cl}(x, y, t) 
+ (V(x) - V(y)) \rho_{\cl}(x, y, t)
\nonumber \\
&+& \frac{\hb^2}{2m} \left [ \frac{1}{ | \rho_{\cl}(x, y, t) |} \left( \frac{\pa^2}{ \pa x^2 } - \frac{\pa^2}{ \pa y^2 }   \right) | \rho_{\cl}(x, y, t) | \right ] \rho_{\cl}(x, y, t)  ,
\end{eqnarray}
and then the corresponding {\it nonlinear} transition wave equation 
\begin{eqnarray} \label{eq: tran-vN}
i\hb \frac{\pa}{\pa t} \rho_{\ep}(x, y, t) &=& - \frac{\hb^2}{2m} 
\left( \frac{\pa^2}{ \pa x^2 } - \frac{\pa^2}{ \pa y^2 }   \right) \rho_{\ep}(x, y, t) 
+ (V(x) - V(y)) \rho_{\ep}(x, y, t)
\nonumber \\
&+& (1-\ep) \frac{\hb^2}{2m} \left [ \frac{1}{ | \rho_{\ep}(x, y, t) |} \left( \frac{\pa^2}{ \pa x^2 } - \frac{\pa^2}{ \pa y^2 }   \right) | \rho_{\ep}(x, y, t) |  \right ] \rho_{\ep}(x, y, t)    ,
\end{eqnarray}
for a smooth transition between the two extreme regimes. This equation contains the transition parameter $\ep$ ranging from one (the quantum regime) to zero (the classical regime). In this equation $ \rho_{\ep}(x, y, t) $ is the density matrix in the transition regime. 
The classical density matrix $ \rho_{\cl}(x, y, t) $ has only a mathematical significance \cite{Holland-book-1993}.
We have then proved that this nonlinear differential equation in the range $ 0 < \ep \leq 1 $ is equivalent to the linear one according to 
\begin{eqnarray} \label{eq: scaled-vN}
i\thb \frac{\pa}{\pa t} \ti{\rho}(x, y, t) &=& - \frac{\thb^2}{2m} 
\left( \frac{\pa^2}{ \pa x^2 } - \frac{\pa^2}{ \pa y^2 }   \right) \ti{\rho}(x, y, t) 
+ (V(x) - V(y)) \ti{\rho}(x, y, t)   ,
\end{eqnarray}
where $ \thb = \hb \sqrt{\ep} $ is known as the scaled Planck constant and the scaled density matrix $\trho(x, y, t)$ is related to the transition one through
\begin{eqnarray} \label{eq: trho-tranrho}
\trho(x, y, t) &=& \rho_{\ep}(x, y, t) \exp\left[ i \frac{ S_{\ep}(x, y, t) }{ \hb } \left( \frac{1}{ \sqrt{\ep} }- 1 \right) \right]  ,
\end{eqnarray}
$ S_{\ep} $ being the phase of the transition density matrix $ \rho_{\ep}(x, t) $. Using the polar form $\rho_{\ep}(x, y, t) = A_{\ep}(x, y, t) e^{i S_{\ep}(x, y, t) / \hb} $, then $\trho(x, y, t) = A_{\ep}(x, y, t) e^{i S_{\ep}(x, y, t) / \thb} $ for the scaled density matrix in its polar form.
Note that the classical regime is a singularity and cannot be described by this scaled equation.
%
%Comparison of Eqs. (\ref{eq: tran-vN}) and \eqref{eq: scaled-vN} reveals% 
Note that for the classical density matrix one has that $ \rho_{\cl}(x, t) = \rho_{\ep=0}(x, y, t) = A_{\ep=0}(x, y, t) e^{S_{\ep=0}(x, y, t) / \hb} $. 

By scaling both the space and time coordinates, one can easily see that the scaled von Neumann equation \eqref{eq: scaled-vN} is just the standard von Neumann equation. By scaling the time and space as follows
\begin{numcases}~
T = \frac{t}{ \sqrt{\ep} } , \label{eq: sc-time}\\
X = \frac{x}{ \sqrt{\ep} }   , \label{eq: sc-space}
\end{numcases}
Eq. \eqref{eq: scaled-vN} can be written as
\begin{eqnarray}
i \hb \frac{\pa}{\pa T} \trho(X, Y, T) &=& - \frac{\hb^2}{2m} 
\left( \frac{\pa^2}{ \pa X^2 } - \frac{\pa^2}{ \pa Y^2 }   \right) \trho(X, Y, T) + (V(X) - V(Y)) \trho(X, Y, T)   ,
\end{eqnarray}
which is the usual von Neumann equation. Then, one has that $ \trho(x, y, t) \propto \rho(X, Y, T) $, $ \rho $ being the density matrix in the quantum regime. Taking into account the normalization condition, one gets
\begin{eqnarray}
\trho(x, y, t) &=& \frac{1}{ \sqrt{\ep} } \rho(X, Y, T)
\end{eqnarray}
for the relation between the scaled density matrix and the quantum density matrix. Note that in addition to the time and space coordinates, the width and center in the case of a Gaussian wave packet has to be scaled too. 
Following the above argument we could call the theory of the transition regime $ 0 < \ep < 1 $, the scaled quantum mechanics.

Our goal is now to use the formalism of the scaled quantum mechanics to find the amplitude and phase of the scaled density matrix and 
afterwards the corresponding quantities associated with the classical regime. In this way, as mentioned above, both classical and quantum mechanics can be described with the same language and the corresponding transition is established in a uniform and continuous way. 
The scaled von Neumann equation solution is given by
\begin{eqnarray} \label{eq: sc-von-sol}
\trho(x, y, t) &=& \la x | \tU(t) \ti{ \hat{\rho }}_0 \tU^{\dagger}(t)| y \ra
= \int dx' \int dy' \tG(x, x' , t) \tG^*(y, y' , t) \trho_0(x', y')  ,
\end{eqnarray}
where in the second equality we have introduced the scaled propagator as the matrix elements, in the coordinate space, of the {\it scaled} evolution operator $ \tU(t) $. For time-independent potentials, the scaled evolution operator is $ \tU(t) = e^{-i \tH t / \thb } $ with $ \tH $ being the corresponding scaled Hamiltonian. Using then the eigenvalue equation 
\begin{eqnarray} \label{SSE}
\tH | \tu \ra &=& \tE | \tu \ra   ,
\end{eqnarray}
one can then write the scaled propagator as  
\begin{eqnarray} \label{eq: sc-prop}
\tG(x, x', t) &=& \sum_n e^{-i \tE_n t / \thb} \tu_n(x) \tu_n^*(x')  ,
\end{eqnarray}
with $ \tE_n $ and $ \tu_n(x) $ being the eigenvalues and the corresponding eigenfunctions of the scaled Hamiltonian, respectively.
Finally, the solution of Eq. \eqref{eq: sc-von-sol} can be expressed as 
\begin{eqnarray}\label{eq: sc-von-sol2}
\trho(x, y, t) &=& \sum_n \sum_m \tc_{nm} e^{-i ( \tE_n - \tE_m ) t / \thb} ~ \tu_n(x) \tu_m^*(y)  ,
\end{eqnarray}
where
\begin{eqnarray} \label{eq: expan-coeff}
\tc_{nm} &=& \int dx' \int dy' ~ \tu_n^*(x') \tu_m(y') \trho_0(x', y')  ,
\end{eqnarray}
with the normalization condition
\begin{eqnarray} \label{eq: nor-con}
\sum_n \tc_{nn} &=& 1  .
\end{eqnarray}

\section{Propagation of a scaled Gaussian mixed state in simple potentials}

In this section, we study evolution of a scaled Gaussian mixed state in simple potentials: free particle, linear potential and simple harmonic oscillator. We also analyze the case of a hard wall in the origin i.e., free propagation in the half-space $ x \geq 0 $. We first solve the scaled von Neumann equation and  from the corresponding solution, one obtains the classical density matrix using the procedure outlined above.

Consider an ensemble of particles described initially with the Gaussian mixed state \cite{DoAn-LP-2002} 
\begin{eqnarray} \label{eq: rho0}
\rho_0(x, y) &=& \frac{1}{ \sqrt{2\pi} \si_0 }
\exp \left[  
- \frac{1}{2\si_0^2} \left( \frac{x^2+y^2}{2(1-\lam)} - \frac{\lam}{1-\lam} x y - x_0 (x+y) + x_0^2  \right) + i \frac{p_0}{\hb}(x-y)
\right] , \nonumber \\
\end{eqnarray}
where $x_0$ denotes the mean value of the position operator, $\si_0$ the width and the kick momentum is $p_0$. Impurity of the state, $ 1 - \tr\{ \hat{\rho}_0^2 \}$, is quantified by the parameter $\lambda$, being zero for the pure state and any value in the range (0 , 1) when the state is impure.
From Eq. \eqref{eq: trho-tranrho}, one sees that the initial scaled density matrix is given by
\begin{eqnarray} \label{eq: trho0}
\trho_0(x, y) &=& \frac{1}{ \sqrt{2\pi} \si_0 }
\exp \left[  
- \frac{1}{2\si_0^2} \left( \frac{x^2+y^2}{2(1-\lam)} - \frac{\lam}{1-\lam} x y - x_0 (x+y) + x_0^2  \right) + i \frac{p_0}{\thb}(x-y)
\right]  . \nonumber \\
\end{eqnarray}

\subsection{Propagation in free space}

Using the scaled free particle propagator 
\begin{eqnarray} \label{eq: sc-prop-free}
\tG_{\fr}(x, x', t) &=& \sqrt{ \frac{m}{2\pi i \thb t} } \exp \left[ \frac{i m}{2\thb t} (x-x')^2 \right]  
\end{eqnarray}
in Eq. \eqref{eq: sc-von-sol}, one obtains
\begin{eqnarray}
\tA(x, y, t) &=& \frac{1}{ \sqrt{2\pi} \tsi_t }
\exp \left[ 
- \frac{1}{2 \tsi_t^2} \left( \frac{x^2+y^2}{2(1-\lam)} - \frac{\lam}{1-\lam} x y + \frac{2 x_0 p_0 t}{m} - (x_0 + \frac{p_0 t}{m}) (x+y) + x_0^2 + \frac{p_0^2 t^2}{m^2} \right) 
\right] , \nonumber \\    \label{eq: R_free}
 \\
\tS(x, y, t) &=& \frac{\thb^2 t}{8 m \si_0^2 \tsi_t^2} \frac{1+\lam}{1-\lam}[x^2-y^2-2x_0(x-y)] + \frac{\si_0^2}{\tsi_t^2} p_0 (x-y) \label{eq: S_free}  ,
\end{eqnarray}
for the amplitude and phase of the density matrix, $ \trho(x, y, t) = \tA(x, y, t) e^{i \tS(x, y, t) / \thb} $, where 
\begin{eqnarray} \label{eq: sigmat}
\tsi_t &=& \si_0 \sqrt{ 1 + \frac{1+\lam}{1-\lam} \frac{ \thb^2 t^2 }{ 4 m^2 \si_0^4 } }  
\end{eqnarray}
is the rms width $ \sqrt{ \la ( \hat{x} - \la \hat{x} \ra)^2 \ra } $ of the state. 
The scaled density matrix in the momentum representation is obtained by taking the Fourier transform of $ \trho(x, y, t) = \tA(x, y, t) e^{i \tS(x, y, t) / \thb} $,
\begin{eqnarray} 
\tvrho(p, q, t) &=& \la p | \hat{\rho}(t) | q \ra = 
\frac{1}{2\pi \thb} \int_{-\infty}^{\infty} dx \int_{-\infty}^{\infty} dy ~ e^{-i(p x - q y)/\thb} \trho(x, y, t) \nonumber \\
&=&
\frac{\si_0}{\thb} \sqrt{ \frac{2}{\pi} \frac{1-\lam}{1+\lam} }
\exp \left[
-i \frac{ (p-q)[(p+q)t + 2 m x_0] } { 2 m \thb }
\right. \nonumber \\
&-& \left. \frac{\si_0^2 [ (p-p_0)^2 +(q-p_0)^2 - 2 \lam (p-p0)(q-p_0) ] }{ (1+\lam)\thb^2 }
\right] 
\label{eq: varrho}
\end{eqnarray}
where Eqs. \eqref{eq: R_free} and \eqref{eq: S_free} have been used.
The probability density in the position and  momentum representations are then expressed as 
\begin{eqnarray} 
\ti{P}(x, t) &=& \trho(x, x, t) = \frac{1}{ \sqrt{2\pi} \tsi_t } \exp\left[ - \frac{1}{2 \tsi_t^2}  \left( x - (x_0 + \frac{p_0 t}{m}) \right)^2 \right] \label{eq: free-prob-den}   , \\
\tilde{\Pi}(p) &=& \tvrho(p, p, t)  = \frac{1}{ \sqrt{2\pi} \tsi_p }
\exp \left[ - \frac{ (p-p_0)^2 }{ 2 \tsi_p^2 } \right] \label{eq: sc-prob-mom}  ,
\end{eqnarray}
respectively where
\begin{eqnarray} \label{eq: mom-dis-width}
\tsi_p &=& \frac{ \thb }{ 2 \si_0 } \sqrt{ \frac{1+\lam}{1-\lam} }
\end{eqnarray}
displays the width of the scaled distribution function in the momentum space.
As one expects, the momentum space probability density is independent on time.
For the first moments, one has that
\begin{numcases}~
\wti{ \la \hat{x} \ra }(t) = x_0 + \frac{p_0 t}{m} , \\
\wti{ \la \hat{x}^2 \ra }(t) = \left( x_0 + \frac{p_0 t}{m} \right)^2 + \tsi_t^2  , \\
\wti{ \la \hat{p} \ra }(t) = p_0 , \\
\wti{ \la \hat{p}^2 \ra }(t) = p_0^2 + \frac{\thb^2}{4\si_0^2} \frac{1+\lam}{1-\lam}  , \label{eq: sc-p2av}
\end{numcases}
and  the product of uncertainties in position and momentum is written as
\begin{eqnarray}
\wti{\Del x} \wti{\Del p} &=& \frac{\thb}{2} \sqrt{ \frac{1+\lam}{1-\lam} } \sqrt{ 1 + \frac{1+\lam}{1-\lam} \frac{ \thb^2 t^2 }{ 4 m^2 \si_0^4 } }  .
\end{eqnarray}

If Bohmian perspective is extended to the transition regime, one can introduce scaled trajectories. Then, the {actual} momentum distribution function or the probability distribution of particle momentum 
induced by the position distribution $ \ti{P}(x, t) $ is computed through \cite{Holland-book-1993}
\begin{eqnarray} \label{eq: actaul-mom-dis}
\ti{g}(p, t) &=& \int dx^{(0)} \ti{P}(x^{(0)}, 0) \delta( p - \pa_x \ti{S}|_{y=x=\ti{x}(x^{(0)}, t)} ),
\end{eqnarray}
where $ \ti{x}(x^{(0)}, t) $ is the scaled trajectory with the initial condition $ \ti{x}(x^{(0)}, 0) = x^{(0)} $. Note that in this scheme, momentum of particles in all regimes, quantum, transition and classical, is determined by the gradient of the phase and thus classical mechanics is a first order theory NOT the customary second order one. Using (18) the momentum field 
\begin{eqnarray} \label{eq: sc-mom-filed}
\ti{p}(x, t) &=& \pa_x\ti{S}(x, t) = \frac{p_0 \si_0^2}{\ti{\si}_t^2} + \frac{\thb^2 t}{4m\si_0^2 \ti{\si}_t} \frac{1+\lam}{1-\lam} (x -x _0)
\end{eqnarray}
is obtained which by integration yields
\begin{eqnarray} \label{eq: sc-trajs}
\ti{x}(x^{(0)}, t) &=& x_0 + \frac{p_0}{m} t + (x^{(0)}-x_0) \frac{ \ti{\si}_t }{\si_0}
\end{eqnarray}
for the {\it scaled} trajectories.
Now, by using Eqs. \eqref{eq: sc-mom-filed} and \eqref{eq: sc-trajs} in \eqref{eq: mom-dis-sc} and taking the integral with use of Dirac delta function properties, one obtains
\begin{eqnarray} \label{eq: actual-mom-dis}
\ti{g}(p, t) &=& \frac{1}{\sqrt{2\pi} \ti{\Sigma}_t} \exp \left[ - \frac{(p-p_0)^2}{2\ti{\Sigma}_t^2} \right]
\end{eqnarray}
for the actual momentum distribution where
\begin{eqnarray} \label{eq: actaul-mom-width}
\ti{\Sigma}_t &=& m \dot{\ti{\si}}_t = \ep \frac{\hb^2 t}{4m\si_0^2 \ti{\si}_t } \frac{1+\lam}{1-\lam}
\end{eqnarray}
is the actual momentum width. Comparison of \eqref{eq: mom-dis-width} with \eqref{eq: actaul-mom-width} shows that in the limit $ t \to \infty $, $ \ti{\Sigma}_t \to \tsi_p $ revealing that in performing a momentum measurement the actual momentum distribution coincides with that obtained in the measurement.

From Eqs. \eqref{eq: R_free} and \eqref{eq: S_free}, one obtains
\begin{equation}
A_{\cl}(x, y, t) = \frac{1}{ \sqrt{2\pi} \si_0 }
\exp \left[ 
- \frac{1}{2 \si_0^2} \left( \frac{x^2+y^2}{2(1-\lam)} - \frac{\lam}{1-\lam} x y + \frac{2 x_0 p_0 t}{m} - (x_0 + \frac{p_0 t}{m}) (x+y) + x_0^2 + \frac{p_0^2 t^2}{m^2} \right) 
\right]  \label{eq: Rcl_free}
\end{equation}
and
\begin{eqnarray}
S_{\cl}(x, y, t) &=& p_0 (x-y) , \label{eq: Scl_free}
\end{eqnarray}
for the amplitude and phase of the classical density matrix, respectively. From these equations, one has that the classical probability density is independent on the impurity parameter $ \lam $.
\begin{eqnarray} \label{eq: cl-prob-den}
P_{\cl}(x, t) &=& \rho_{\cl}(x, x, t)
= \frac{1}{\sqrt{ 2\pi} \si_0 } \exp \left[ - \frac{1}{2 \si_0^2} \left( x - x_0 - \frac{p_0 t}{m} \right)^2  \right]  .
\end{eqnarray}
This relation shows that the classical density preserves its shape while propagating freely with the constant velocity $p_0/m$.
%, as a whole, a signification of the solitary solution, with the constant velocity $p_0/m$.
%

From Eq. \eqref{eq: Scl_free}, the momentum field is given by $ p = \pa_x S_{\cl}|_{y=x} = p_0 $ and classical trajectories are $ x_{\cl}(t) = x^{(0)} + p_0 t /m $, $x^{(0)}$ being the initial value of the position of the classical particle. Then, for the classical actual momentum distribution function we have that
\begin{eqnarray}
g_{\cl}(p) &=& \int dx^{(0)} P_{\cl}(x^{(0)}, 0) \delta( p - \pa_x S_{\cl}|_{y=x=x_{\cl}(t)} ) = \del(p - p_0)  ,
\end{eqnarray}
showing all classical particles have the same momentum i.e., dispersion in momentum is zero. This result coincides with Eq. \eqref{eq: actual-mom-dis} in the limit $ \ep \to 0 $ i.e., $ \lim_{\ep \to 0} \ti{g}(p) = g_{\cl}(p) $. In fact from Eqs. \eqref{eq: sc-prob-mom} and \eqref{eq: actual-mom-dis} one sees that $ \lim_{\ep \to 0} \ti{\Pi}(p) = \lim_{\ep \to 0} \ti{g}(p) = \del(p - p_0) $ i.e., in the classical limit both momentum distributions become the same.

Let us consider now the evolution of the state \eqref{eq: trho0} in the positive half-space described by the potential 
\begin{equation}
V(x) = 
\begin{cases}
0 , & x >0 \\
\infty,  & x \leq 0   .
\end{cases}
\end{equation}
In this case, the propagator is given by
\begin{eqnarray} \label{eq: sc-prop-hw}
\tG(x, x', t) &=& \tG_{\fr}(x, x', t) - \tG_{\fr}(-x, x', t) .
\end{eqnarray}
Using this relation in Eq. \eqref{eq: sc-von-sol}, one has that
\begin{eqnarray} \label{sc-trho-hw}
\trho(x, y, t) &=& \trho_{\fr}(x, y, t) -  \trho_{\fr}(x, -y, t) - \trho_{\fr}(-x, y, t) + \trho_{\fr}(-x, -y, t)  ,
\end{eqnarray}
for the evolution of the mixed state \eqref{eq: trho0} where the amplitude and the phase of  $ \trho_{\fr}(x, y, t) $ are given by Eqs. \eqref{eq: R_free} and \eqref{eq: S_free}, respectively. 
Note that since we take the initial Gaussian state \eqref{eq: trho0} well localized in a point far from the hard wall, we can extend the domain of integration to the whole space with a negligible error. 

The corresponding equation for the classical regime is
\begin{eqnarray} 
\rho_{\cl}(x, y, t) &=& A_{\cl}(x, y, t) \, e^{i S_{\cl}(x, y, t)/\hb} -  A_{\cl}(x, -y, t) \, e^{i S_{\cl}(x, -y, t)/\hb} - A_{\cl}(-x, y, t) \, e^{i S_{\cl}(-x, y, t)/\hb} \nonumber \\
&+& A_{\cl}(-x, -y, t) \, e^{i S_{\cl}(-x, -y, t)/\hb} 
\label{cl-rho-hw}
\end{eqnarray}
where $A_{\cl}$ and $S_{\cl}$ are given by Eqs. \eqref{eq: Rcl_free} and \eqref{eq: Scl_free}, respectively.

\begin{figure}  
\includegraphics[width=12cm,angle=-0]{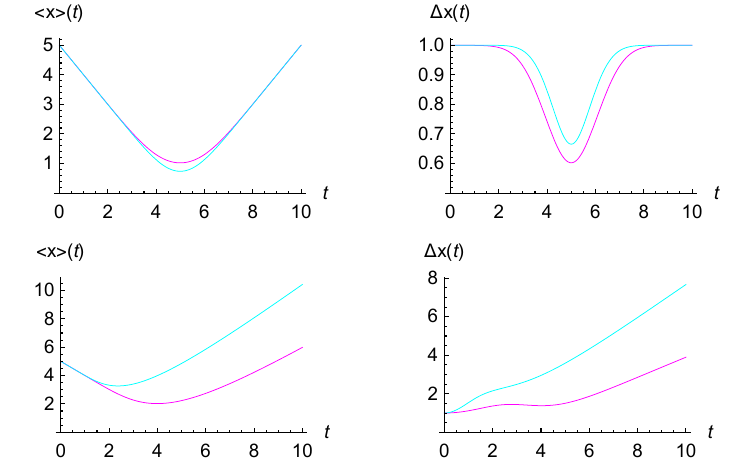}
\caption{
The position expectation value (left column) and dispersion in position (right column) for the classical (top row) and quantum regimes (bottom row) in the free evolution of the state \eqref{eq: rho0} in the positive half-space, for $ \lam = 0 $ (magenta curves) and $ \lam = 0.7 $ (cyan curves).
For numerical calculations, we have used $m = 1$, $\hb =1$, $\si_0 = 1$, $ x_0 = 5 $ and $p_0=-1$.}
\label{fig: clqm-xav-delx}
\end{figure}

In Figure \ref{fig: clqm-xav-delx}, we have plotted the position expectation value (left column) and the dispersion in position (right column) for the classical regime (top row) and quantum regime (bottom row) for the free evolution of the state \eqref{eq: rho0} in the positive half-space for $ \lam = 0 $ (magenta curves) and $ \lam = 0.7 $ (cyan curves). For numerical calculations, we have used $m = 1$, $\hb =1$, $\si_0 = 1$, $ x_0 = 5 $ and $p_0=-1$. This figure shows that quantum particles, in average, reflect from the hard wall at a far greater distance and also in a shorter time in comparison to the classical particles. This can be explained by the quantum effective force. Furthermore, the impurity is determinant in this reflection from the wall; while the reflection time is classically independent of impurity, quantum mechanically, impurity makes the time of reflection shorter.

%===============================================
%Figure 
%
%
%===============================================

\subsection{Propagation in the uniform gravitational field}

In this section, we consider evolution of the mixed state \eqref{eq: trho0} in the linear potential
\begin{eqnarray} \label{eq: lin-pot}
V(x) &=& m g x  ,
\end{eqnarray}
which describes a uniform gravitational field with strength $g$. Using the corresponding scaled propagator
\begin{eqnarray} \label{eq: sc-prop-lin}
\tG(x, x', t) &=& \sqrt{ \frac{m}{2\pi i \thb t} } \exp \left[ \frac{i m}{2\thb t} (x-x')^2 
- i \frac{m g}{2\thb}(x+x')t - i \frac{m g^2}{24\thb} t^3
\right]  ,
\end{eqnarray}
and taking the kick momentum $p_0 = 0$, one reaches 
\begin{eqnarray} 
\ti{P}(x, t) &=& \frac{1}{ \sqrt{2\pi} \tsi_t } \exp\left[ - \frac{1}{2 \tsi_t^2}  \left( x - (x_0- \frac{g t^2}{2}) \right)^2 \right] \label{eq: lin-prob-den} , \\
\ti{J}(x, t) &=& 
\frac{ \{ 2 \thb^2 (1+\lam) (x - x_0) - g [ \thb^2 t^2 (1+\lam) + 8 m^2 \si_0^4 (1-\lam) ] \} t  }{ 2 [ \thb^2 t^2 (1+\lam) + 4 m^2 \si_0^4 (1-\lam) ] }
 \ti{P}(x, t) , \label{eq: lin-pcd}
\end{eqnarray}
for the probability density (PD) and the probability current density (PCD) respectively, with $\tsi_t$ being the rms width of PD given by Eq. \eqref{eq: sigmat}. 
After the scaled velocity field is obtained by the ratio of the PCD to PD, one can express the  scaled trajectories as
\begin{eqnarray} \label{eq: lin-sc-trajs}
\ti{x}(t, x^{(0)}) &=& \left( x_0 - \frac{g t^2}{2} \right) + ( x^{(0)} - x_0) \frac{\tsi_t}{\si_0}  ,
\end{eqnarray}
showing the typical dressing scheme \cite{NaMi-book-2017}.
For the classical density matrix, $ \rho_{\cl}(x, y, t) = A_{\cl}(x, y, t) e^{ i S_{\cl}(x, y, t) / \hb} $ the corresponding expressions are given by 
\begin{eqnarray}
&&A_{\cl}(x, y, t) = 
\nonumber \\
&&\frac{1}{\sqrt{2\pi} \si_0} \exp \left[  
-\frac{ 2 (x^2 + y^2) - 4 \lam x y - 4(1 - \lam) x_0 (x + y - x_0) +  g t^2  (1 - \lam) [2 (x + y) - 4 x_0 + g t^2] }{8 (1-\lam) \si_0^2} 
\right]  \nonumber \\  \label{eq: lin-cl-R} 
\end{eqnarray}
and
\begin{eqnarray}
S_{\cl}(x, y, t) &=& - m g t (x - y)  ,	\label{eq: lin-cl-S}
\end{eqnarray}
which leads to
\begin{eqnarray}
P_{\cl}(x, t) &=& \frac{1}{\sqrt{2\pi} \si_0} \exp\left[ - \frac{1}{2 \si_0^2}  \left( x - (x_0- \frac{g t^2}{2}) \right)^2 \right] , \label{eq: lin-cl-Prob}
\end{eqnarray}
for the classical probability density and
\begin{eqnarray} \label{eq: lin-cl-trajs}
x_{\cl}(t, x^{(0)}) &=& x^{(0)} - \frac{1}{2} g t^2  ,
\end{eqnarray}
which is just the limit $ \ep \to 0 $ of the scaled trajectories, Eq. \eqref{eq: lin-sc-trajs}. This shows that all classical particles in the ensemble move with the same velocity $ - g t $; a result which can be confirmed by computing the momentum distribution of the classical ensemble of particles. Note that for the scaled motion, particles move with different velocities as can be seen by the time derivative of Eq. \eqref{eq: lin-sc-trajs}, which shows that the velocity depends on the initial position of particle.

\subsection{Propagation in a simple harmonic potential}

In this section, the evolution of the mixed state \eqref{eq: trho0} is studied for the harmonic potential
\begin{eqnarray} \label{eq: har-pot}
V(x) &=& \frac{1}{2} m \om^2 x^2  .
\end{eqnarray}
Since the relation for the density matrix is quite cumbersome, we only give the probability density and the probability current density when the kick momentum is zero, $p_0 = 0$. By using the scaled propagator
\begin{eqnarray} \label{eq: SHO-prop}
\ti{G}(x, x', t) &=& \sqrt{ \frac{ m\om }{2 \pi i \thb \sin(\om t) } }
\exp \left[ i \frac{ m\om }{2 \thb \sin(\om t) } \{ (x^2 + x'^2)\cos(\om t) - 2 x x' \} \right]  ,
\end{eqnarray}
the solution of the scaled von Neumann equation is
\begin{eqnarray} \label{eq: har-probden}
\ti{P}(x, t) &=& \frac{1}{ \sqrt{2\pi} \tsi_t } \exp\left[ - \frac{ (x-x_0 \cos(\om t))^2 }{2 \tsi_t^2} \right]
\end{eqnarray}
for PD where
\begin{eqnarray} \label{eq: har-width}
\tsi_t &=& \si_0 \sqrt{ \cos^2(\om t) + \frac{1+\lam}{1-\lam} \frac{\hb^2}{4m^2\si_0^4} \frac{ \sin^2(\om t)}{\om^2}}
\end{eqnarray}
is its rms width. For the scaled PCD, one has that
\begin{eqnarray} \label{eq: har-pcd}
\ti{J}(x, t) &=& \frac{ 2 \om  \sin (\om t ) [\thb^2 (1+\lam) (x \cos (\om t )- x_0)- 4 x (1-\lam ) m^2 \si_0^4 \om^2 \cos (\om t )] }
{ 2 \thb^2 (1+\lambda) \sin^2(\om t) + 8 (1-\lambda ) m^2  \om^2 \si_0^4 \cos^2(\om t) }
 \ti{P}(x, t) .
\end{eqnarray}
Then, the scaled velocity field can be obtained by the ratio of PCD to PD from which one obtains scaled trajectories expressed as
\begin{eqnarray} \label{eq: har-sc-trajs}
\ti{x}(t, x^{(0)}) &=& x_0 \cos(\om t) + (x^{(0)} - x_0) \frac{\tsi_t}{\si_0}  .
\end{eqnarray}

From the amplitude and phase of the transition density matrix, not given here, one has
\begin{eqnarray}   \label{eq: har-cl-R}
&&A_{\cl}(x, y, t) = 
\nonumber \\
&&\frac{1}{\sqrt{2\pi} \si_0 |\cos(\om t)|} \exp\left[  
- \frac{ x^2 + y^2 - 2 \lam x y - 2 x_0 (1-\lam) \cos(\om t)~(x+y) + 2 x_0^2 (1-\lam)\cos^2(\om t) }{ 4(1-\lam) \si_0^2 \cos^2(\om t) } \right] \nonumber \\
\end{eqnarray}
and
\begin{eqnarray} \label{eq: har-cl-S}
S_{\cl}(x, y, t) &=& - \frac{ m \om (x^2-y^2) }{2} \tan(\om t)	, 
\end{eqnarray}
for the amplitude and phase of the classical density matrix, respectively. One then sees that the classical density matrix $ \rho_{\cl}(x, y, t) = A_{\cl}(x, y, t) e^{i S_{\cl}(x, y, t) / \hb} $ fulfils the classical von Neumann equation. From Eq. \eqref{eq: har-cl-S}, the classical momentum field is given by $ p_{\cl}(x, t) = \pa_x S(x, y, t) |_{y=x} = - m \om \tan(\om t) x $ and the classical trajectories are expressed as
\begin{eqnarray} \label{eq: har-cl-trajs}
x_{\cl}(t, x^{(0)}) &=& x^{(0)} | \cos(\om t) |.
\end{eqnarray}
%
%which itself is the limit $ \ep \to 0 $ of \eqref{eq: har-sc-trajs}.

\section{Scaled bouncing ball} \label{sec: SBB}

Let us consider now a ball of mass $ m $ which bounces on a hard wall in the presence of 
(i) a gravitational field with strength $ g $ and (ii) a harmonic oscillator  with frequency $ \om $. 
Our BB is initially described by the Gaussian wave packet
\begin{eqnarray} \label{eq: psi0G}
\psi_0(z) &=& \frac{1}{(2\pi \si_0^2)^{1/4}} \exp \left[ - \frac{(z-z_0)^2}{4\si_0^2} \right]   ,
\end{eqnarray}
$ \si_0 $ and $ z_0 $ being  the width and centre of the packet, respectively. In this section we prefer to represent the position variable with $z$ instead of $x$.
Note that this state is just the state \eqref{eq: rho0} with the impurity parameter $ \lam = 0 $ and zero 
kick momentum. So, since we are dealing with a pure state, one can use the Schr\"odinger formalism 
instead of the von Neumann one in order to study its time evolution.

\subsection{Gravitational field}

In this case, the potential energy is given by
\begin{eqnarray} \label{eq: gbb-pot}
V(z) &=& 
\begin{cases}
m g z , & z > 0 \\
+\infty , & z \leq 0   .
\end{cases}
\end{eqnarray}
%
%
%\subsubsection{Quantum gravitational bouncing ball}
%
Let us first revise  the standard quantum regime i.e., the standard Schr\"odinger equation 
\begin{eqnarray} \label{eq: TISE}
\left( \frac{\hb^2}{2m} \frac{d^2}{d z^2} + m g z \right) u_n(z) &=& E_n u_n(z), \qquad z > 0
\end{eqnarray}
which leads to the eigenvalues and eigenfunctions expressed as \cite{VaSo-book-2004}
\begin{numcases}~
E_n = - R_n \left( \frac{1}{2} m g^2 \hb^2 \right)^{1/3}, \qquad n = 1, 2, 3, \cdots\label{eq: En-exact} \\
u_n(z) = \sqrt{\be} \frac{1}{\Ai'(R_n)} \Ai(\be z + R_n)   , \qquad \be = \left( \frac{2 m^2 g}{\hb^2} \right)^{1/3} , \label{eq: eigenfunc}
\end{numcases}
where $\Ai(z)$ and $\Ai'(z)$ are the Airy function and its derivative, respectively. Here, $R_n$ is the 
n$^{\text{th}}$ root of the Airy function i.e., $ \Ai(R_n) = 0 $. It is worth mentioning that the quantization 
condition in the WKB approximation gives
\begin{eqnarray} \label{eq: En-WKB}
E_n^{\text{WKB}} &=&  \left[ \frac{3\pi}{2} \left( n-\frac{1}{4} \right) \right]^{2/3}  \left( \frac{1}{2} m g^2 \hb^2 \right)^{1/3}
\end{eqnarray}
for the energy levels \cite{Sakurai-book-1994}. The WKB approximation gives the correct ground-state 
energy up to only $0.8 \%$. The agreement between the exact result (\ref{eq: En-exact}) and the WKB result 
(\ref{eq: En-WKB}) is obviously much better at higher and higher levels. 

From the evolution of the Gaussian wave packet \eqref{eq: psi0G}, the wave function can be written as
\begin{eqnarray*}
\psi(z, t) &=& \sum_n C_n u_n(z) e^{-i E_n t/\hb}   ,
\end{eqnarray*}
with eigenvalues and eigenfunction given by Eqs. \eqref{eq: En-exact} and \eqref{eq: eigenfunc} respectively, 
and the expansion coefficients are obtained according to
\begin{eqnarray} \label{eq: expan-bb}
C_n &=& \frac{ \sqrt{\be} }{\Ai'(R_n)} \frac{1}{(2\pi \si_0^2)^{1/4}} \int_0^{\infty} dz ~ \Ai(\be z + R_n) \exp \left[ - \frac{(z-z_0)^2}{4\si_0^2} \right]   .
\end{eqnarray}
In the limit $ \si_0 \ll z_0 $ i.e., when the Gaussian wave packet is well localized away from the boundary,
the lower limit of the integral can be replaced by $-\infty$ and the coefficients are  then \cite{Va-AJP-2000}
\begin{eqnarray} \label{eq: expan-bb-approx}
C_n &\approx & \frac{ (2\pi)^{1/4} \sqrt{2 \be \si_0} }{\Ai'(R_n)}  \Ai(\be z_0+R_n+ \be^4 \si_0^4)
\exp\left[ \be^2 \si_0^2 \left( \be z_0 + R_n + \frac{2}{3} \be^4 \si_0^4 \right) \right]   .
\end{eqnarray}
%

%\subsubsection{Scaled quantum gravitational bouncing ball}

After this short discussion of the BB in the quantum regime, let us consider now the scaled theory.
The spectrum of the scaled quantum gravitational BB is obtained from the scaled time-independent Schr\"odinger equation
\begin{eqnarray} \label{eq: sTISE}
\left( - \frac{\thb^2}{2m} \frac{d^2}{d z^2} + mgz \right) \tu_n(z) &=& \tE_n \tu_n(z), \qquad z>0  .
\end{eqnarray}
Comparison to Eq. (\ref{eq: TISE}) reveals that
\begin{numcases}~
\tE_n = - R_n \left( \frac{1}{2} m g^2 \thb^2 \right)^{1/3}   , \label{eq: scEn} \\
\tu_n(z) = \sqrt{\tbe} \frac{1}{\Ai'(R_n)} \Ai(\tbe z + R_n)   , \qquad \tbe = \left( \frac{2 m^2 g}{\thb^2} \right)^{1/3} . \label{eq: scun}
\end{numcases}
Now, according to Eq. (\ref{eq: scEn}), the difference between two adjacent energy eigenvalues is
\begin{eqnarray}
\Delta \tE &=& \tE_{n+1} - \tE_n = \ep^{1/3} \left( \frac{1}{2} m g^2 \hb^2 \right)^{1/3} (R_n-R_{n+1})  ,
\end{eqnarray}
which becomes vanishingly small in the limit $ \ep \to 0 $ confirming a continuum spectrum in the 
classical limit.

In the following, for convenience, dimensionless quantities are going to be used. To this end, 
length, time and energy  are multiplied by $ (\hb^2/2m^2g)^{1/3} $, $ (2\hb/mg^2)^{1/3} $ and 
$ (mg^2 \hb^2/2)^{1/3} $ respectively i.e.,
\begin{numcases}~
z \to (\hb^2/2m^2g)^{1/3} ~ z , \\
t \to ( 2\hb/m g^2  )^{1/3} ~ t , \\
\tE \to ( mg^2 \hb^2/2)^{1/3} ~ \tE    .
\end{numcases}
Thus,  the corresponding eigenfunctions and wave functions are given in terms of 
$ (2m^2g/\hb^2)^{1/6} $ and the
propagator and the probability density as  $ (2m^2g/\hb^2)^{1/3} $. By substituing Eqs. (\ref{eq: scEn}) 
and (\ref{eq: scun}) into Eq. (\ref{eq: sc-prop}) one has that
\begin{eqnarray} \label{eq: scG1}
\ti{G}(z, z', t) &=& \frac{1}{ \sqrt[3]{\ep} }
\sum_{n=1}^{\infty} \frac{ \exp[ i R_n t /\sqrt[6]{\ep} ]} { (\Ai'(R_n))^2 } 
~ \Ai \left( \frac{z}{\sqrt[3]{\ep}} + R_n \right) ~ \Ai \left( \frac{z'}{\sqrt[3]{\ep}} + R_n \right)   ,
\end{eqnarray}
from which the scaled time-dependent wave function is expressed as
\begin{eqnarray} 
\tpsi(z, t) &=& \int dz' \ti{G}(z, z', t) \tpsi(z', 0) \label{eq: tpsit}
\\
&=& 
\sum_{n=1}^{\infty} \frac{ \exp[ i R_n t /\sqrt[6]{\ep} ]}{ (\Ai'(R_n))^2 }
~ \Ai \left( \frac{z}{\sqrt[3]{\ep}} + R_n \right)
\int_0^{\infty} dz' ~  \Ai(z'+R_n) ~ \tpsi(\sqrt[3]{\ep}~z', 0)  . \label{eq: tpsit1}
\end{eqnarray}
Let us start with the Gaussian wave function (\ref{eq: psi0G}) which is independent on $ \thb $ and thus 
on $\ep$. This means that  in all regimes, including the classical one, the initial position distribution function 
is $ |\psi_0(z) |^2 $. From Eq. (\ref{eq: tpsit1}), we have that
\begin{eqnarray} 
\tpsi(z, t) &=& \frac{1}{\sqrt[6]{\ep}}
\sum_{n=1}^{\infty} \frac{ e^{ i R_n \bar{t} } }{ (\Ai'(R_n))^2 }
~ \Ai (\bar{z} + R_n ) 
\nonumber \\
& \times &
\frac{1}{ (2\pi \bsi_0^2)^{1/4} }
\int_0^{\infty} dz' ~  \Ai(z'+R_n) ~ \exp \left[ - \frac{(z'-\bar{z}_0)^2}{4\bar{\si}_0^2} \right] 
\label{eq: tpsiG} \\
& \equiv & \frac{1}{\sqrt[6]{\ep}} \bar{\psi}_{\QM}(z, t)    , \label{eq: tpsiG-tpsiGq}
\end{eqnarray}
where we have introduced the bar quantities
\begin{numcases}~
\bar{t} = t / \sqrt[6]{\ep} , \\
\bar{z} = z / \sqrt[3]{\ep} , \\
\bar{\si}_0 = \si_0 / \sqrt[3]{\ep} , \\
\bar{z}_0 = z_0 / \sqrt[3]{\ep}   ,
\end{numcases}
and $ \bar{\psi}_{\QM}(z, t) $ is the wave function in the quantum regime in terms of these bar quantities i.e.,
\begin{eqnarray}
\bar{\psi}_{\QM}(z, t) = \psi_{\QM}(\bar{z},\bar{t})\bigg|_{\si_0 \to \bar{\si}_0, z_0 \to \bar{z}_0}   .
\end{eqnarray}
Therefore, Eq. (\ref{eq: tpsiG-tpsiGq}) shows that the scaled wave function is just the {\it scaled} 
quantum wave function with the scaled time and the scaled length quantities (space coordinate; and 
center and with of the initial Gaussian wave packet).
From this relation, one can now find relations between other scaled and quantum quantities. 
For instance, for the position operator expectation value one has that
\begin{eqnarray} \label{eq: zav}
\ti{\la z \ra}(t) &=& \la \tpsi(t) | z | \tpsi(t) \ra  =  \sqrt[3]{\ep} ~ \la \bar{\psi}_{\QM}(t) | z | \bar{\psi}_{\QM}(t) \ra .
\end{eqnarray}
%
%or for the uncertainty, one obtains
%
%\begin{eqnarray} \label{eq: uncer}
%\wti{\Delta z}(t) &=& \sqrt[3]{\ep} ~ \sqrt{ \la \bar{\psi}_{\QM}(t) | z^2 | \bar{\psi}_{\QM}(t) \ra 
%-  \la \bar{\psi}_{\QM}(t) | z | \bar{\psi}_{\QM}(t) \ra^2
%}
%\end{eqnarray}
%

\begin{figure}  
\includegraphics[width=10cm]{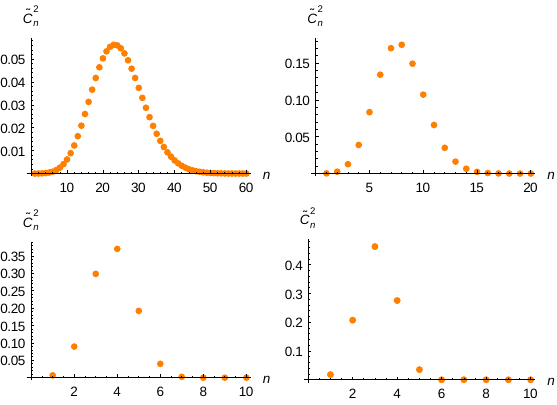}
\caption{
The first scaled expansion weights of the Gaussian wave packet in terms of the scaled eigenstates for the scaled gravitational BB for $ \ep = 0.01 $ (left top panel), $ \ep = 0.1 $ (right top panel), $ \ep = 0.5 $ (left bottom panel) and 
$ \ep = 0.8 $ (right bottom panel).
For numerical calculations, we have used $\si_0 = 1$ and $ z_0 = 5 $ in units of $ (\hb^2/2m^2g)^{1/3} $.}
\label{fig: plotexpansion}
\end{figure}

\begin{figure}  
\includegraphics[width=12cm,angle=0]{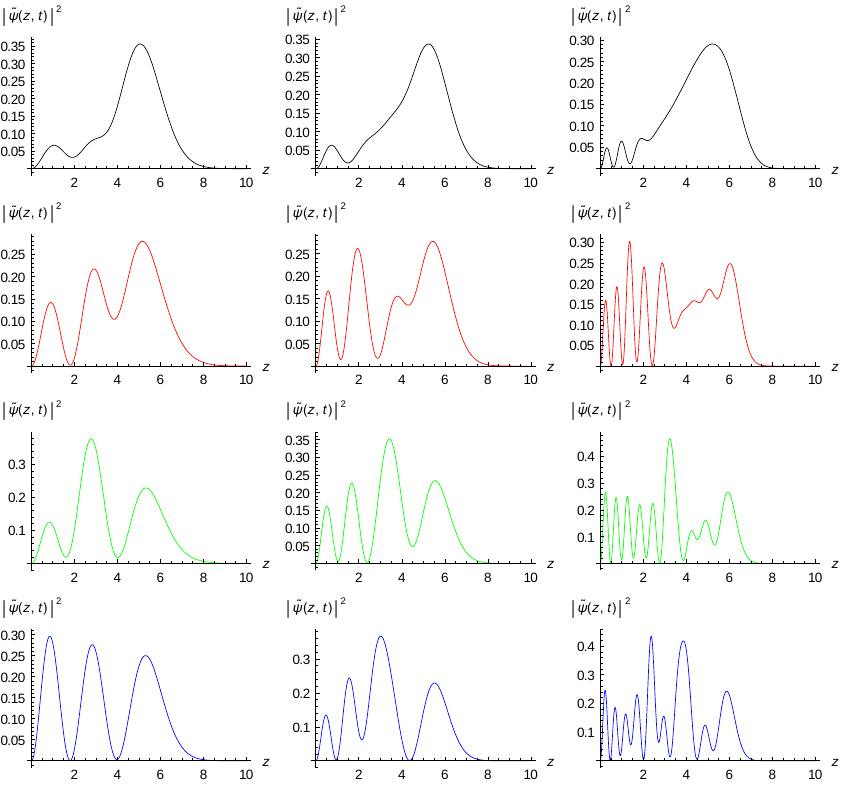}
\caption{
The scaled probability density $ |\tpsi(z, t)|^2 $ for the scaled gravitational BB for $ \ep = 0.1 $ (left column), $ \ep = 0.5 $ (middle column) and $ \ep = 1 $ (right column) at different times, in units of $ (2\hb/mg^2)^{1/3} $, $ t = 5 $ (top row), $ t = 10 $ (second row), $ t = 15 $ (third row) and $ t = 20 $ (bottom row). For numerical calculations, we have used $\si_0 = 1$ and $ z_0 = 5 $ in units of $ (\hb^2/2m^2g)^{1/3} $.}
\label{fig: psi2}
\end{figure}

As Fig. \ref{fig: plotexpansion} shows, when going from the quantum to classical regime more 
eigenfunctions are involved to represent the Gaussian wave packet. 
This is so since in approaching  the classical regime, eigenfunctions become  narrower and narrower 
and thus, in this limit, more and more eigenfunctions are needed to represent the same wave packet \eqref{eq: psi0G}.
In Fig. \ref{fig: psi2}, we have plotted
the scaled probability density versus space coordinate at different times for three different dynamical regimes: 
$ \ep = 0.1 $ (left panel) and $ \ep = 0.5 $ (middle panel) and the quantum regime $ \ep = 1 $ (right panel).
As expected, according to this figure, in this transition and at a given time, oscillations increase and the 
width decrease.

%===============================================
%Figure 
%
%
%===============================================

%===============================================
%Figure 
%
%
%===============================================

\begin{figure}  
\includegraphics[width=12cm,angle=-0]{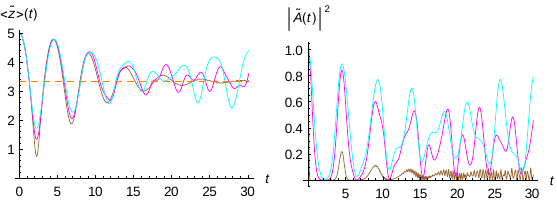}
\caption{
The scaled position expectation value $ \ti{\la z \ra}(t) $, in units of $ (\hb^2/2m^2g)^{1/3} $, (left panel) and the squared modulus of the autocorrelation function, $ |\ti{A}(t)|^2 $, versus time, in units of $ (2\hb/mg^2)^{1/3} $, for the scaled gravitational BB for $ \ep = 0.01 $ (brown curves), $ \ep = 0.5 $ 
(magenta curves) and $ \ep = 1 $ (cyan curves). Orange dashed line on the left panel shows 
$ \la ~ \overline{z_{\cl}(t, z^{(0)})} ~ \ra $.
For numerical calculations,  we have again used $\si_0 = 1$ and $ z_0 = 5 $ in units of $ (\hb^2/2m^2g)^{1/3} $.}
\label{fig: zavautocor}
\end{figure}

On the left panel of Fig. \ref{fig: zavautocor}, we have plotted the expectation value of the position 
operator for three different dynamical  regimes at short times i.e, after six bounces. 
If a classical particle falls at the height $ z^{(0)} $, it collides with the floor at time $ \sqrt{2 z^{(0)} /g} $. 
Assuming an elastic collision, it then bounces off from the wall with velocity $ \sqrt{2 g z^{(0)}} $. 
Motion is classically periodic with a period $ \tau = 2\sqrt{2 z^{(0)} /g} $. Classical trajectory in a 
period of motion is given by
\begin{eqnarray} \label{eq: cl-traj}
z_{\cl}(t, z^{(0)}) &=&
\begin{cases}
-\frac{1}{2} g t^2 + z^{(0)} &  0 \leq t < \tau/2 \\
\\
-\frac{1}{2} g (t- \frac{\tau}{2})^2 + \sqrt{2g z^{(0)}} (t- \frac{\tau}{2})  &  \tau/2 \leq t \leq \tau   .
\end{cases}
\end{eqnarray}
Then, the {\it time} average of $ z_{\cl}(t) $ yields
\begin{eqnarray} \label{eq: time-av}
\overline{z_{\cl}(t, z^{(0)})} &=& \frac{1}{\tau} \int_0^{\tau} dt' z_{\cl}(t') = \frac{2}{3} z^{(0)} .
\end{eqnarray}
Now if we consider an ensemble of classical particles with the initial distribution $ |\psi_0( z^{(0)} )|^2 $ 
where $ \psi_0 $ is the Gaussian wave packet (\ref{eq: psi0G}) then the ensemble average of
$ \overline{z_{\cl}(t, z^{(0)})} $ leads to
\begin{eqnarray} \label{eq: ensem-av}
\la ~ \overline{z_{\cl}(t, z^{(0)})} ~ \ra &=& 
\int  dz^{(0)} ~ \overline{z_{\cl}(t, z^{(0)})} ~ |\psi_0( z^{(0)} )|^2 =  \frac{2}{3} z_0  ,
\end{eqnarray}
$z_0$ being the center of the initial Gaussian wave packet.
As the left panel of Fig. \ref{fig: zavautocor} shows, oscillations are around the classical mean value $ 2 z_0 / 3 = 3.33 $ \cite{FeBeFe-EJP-2019}. 
Revivals of oscillations, not shown in the figure, are seen at longer times. 
On the right panel of this figure, it is depicted the squared modulus of the 
autocorrelation function which itself is defined as \cite{Ro-PR-2004}
\begin{eqnarray} \label{eq: authocor}
\ti{A}(t) &=& \la \tpsi(t) | \psi(0) \ra = \sum_n |\tC_n|^2~ e^{i \tE_n t/\thb} ,
\end{eqnarray}
quantifying the overlap between the initial and evolved states. According to Fig. \ref{fig: zavautocor}, 
when approaching the classical regime, particles bounce off the wall in a closer distance to the wall, 
on average; and the autocorrelation is smaller for dynamical regimes near the classical one.

%===============================================
%Figure 
%
%
%===============================================

\subsection{Harmonic potential}

In this case, the potential energy is 
\begin{eqnarray} \label{eq: hbb-pot}
V(z) &=& 
\begin{cases}
\frac{1}{2} m \om z^2  , & z > 0 \\
+\infty , & z \leq 0   .
\end{cases}
\end{eqnarray}
Because of the hard wall at $z=0$, the solution of the  scaled time-dependent Schr\"odinger equation 
\begin{eqnarray} \label{eq: Scaled Sch}
i \thb \frac{\pa}{\pa t} \tpsi(z, t) &=& \left[ -\frac{\thb^2}{2m} \frac{\pa^2}{\pa z^2} + V(z) \right] \tpsi(z, t) ,
\end{eqnarray}
is given by
\begin{eqnarray} \label{eq: hardwall}
\tPsi(z, t) &=& \tpsi(z, t) - \tpsi(-z, t), \qquad z>0   .
\end{eqnarray}

The evolution of the scaled Gaussian wave packet \eqref{eq: psi0G} under the harmonic oscillator potential 
$ V(z) = m \om^2 z^2 / 2 $ is obtained as
\begin{eqnarray} \label{eq: psiG-sHO}
\tpsi(z, t) &=& \frac{1}{(2\pi \si_0^2)^{1/4}} \sqrt{ \frac{ m\om }{2 \pi i \thb \sin(\om t) } }
\sqrt{\frac{\pi}{\ti{a}_2(t)}} \exp\left[ \ti{a}_0(z, t) + \frac{\ti{a}_1(z, t)^2}{4 \ti{a}_2(t)} \right]   ,
\end{eqnarray}
where 
\begin{numcases}~
\ti{a}_0(z, t) = - \frac{z_0^2}{4\si_0^2} + i \frac{ m \om \cot(\om t) }{ 2\thb }z^2 
\\
\ti{a}_1(z, t) = \frac{z_0}{4\si_0^2} - i \frac{ m \om \csc(\om t) }{ \thb }z 
\\
\ti{a}_2(t) = \frac{1}{4\si_0^2} - i \frac{ m \om \cot(\om t) }{ 2\thb }  ,
\end{numcases}
and Eq. \eqref{eq: tpsit} with the propagator \eqref{eq: SHO-prop} have been used.

A bouncing ball which is initially in the state \eqref{eq: psi0G}, it is described afterwards by 
Eq. \eqref{eq: hardwall} where $ \tpsi(z, t) $ is given by Eq. \eqref{eq: psiG-sHO}. The expectation value of the 
position operator is then 
\begin{eqnarray} 
\wti{\la z \ra}(t) &=& \si_0^2 \cos^2(\om t) \left( 1 + \frac{\thb^2}{ 4 m^2 \si_0^4 \om^2 } \tan^2(\om t) \right)
\exp \bigg[ - \frac{ 2 m^2 z_0^2 \si_0^2 \om^2 }{ 4m^2 \si_0^4 \om^2 + \thb^2 \tan^2(\om t) } \bigg]
\nonumber \\
&\times&
\bigg\{ 
\frac{ 2 \thb m z_0 \om \csc(\om t) }{ \thb^2 + 4 m^2 \si_0^4 \om^2 \cot^2(\om t) }
\exp \bigg[ - \frac{\thb^2 z_0^2}{ 2\thb^2 \si_0^2 + 8 m^2 \si_0^6 \om^2 \cot^2(\om t) } \bigg]
\erfi \bigg[ \frac{ \thb z_0 | \csc(\om t) |~ \sin(\om t)  }{ \si_0 \sqrt{ 2 \thb^2 + 8 m^2 \si_0^4 \om^2 \cot^2(\om t) } } \bigg]
\nonumber \\
& & +
\frac{ 4 m^2 z_0 \si_0^2 \om^2 \cot(\om t) \csc(\om t) }{ \thb^2 + 4 m^2 \si_0^4 \om^2 \cot^2(\om t) }
\exp \bigg[ \frac{ 2m^2 z_0^2 \si_0^2 \om^2 }{ 4 m^2 \si_0^4 \om^2 + \thb^2 \tan^2(\om t) } \bigg]
\erf \bigg[ \frac{ \sqrt{2} m z_0 \si_0 \om | \csc(\om t) |~ \cos(\om t)  }{ \sqrt{ \thb^2 + 4 m^2 \si_0^4 \om^2 \cot^2(\om t) } } \bigg]
 \bigg\}     .  \nonumber \\  \label{eq: zavHBBgaus}
\end{eqnarray}
%

%===============================================
%Figure 
%
\begin{figure}  
\includegraphics[width=12cm,angle=-0]{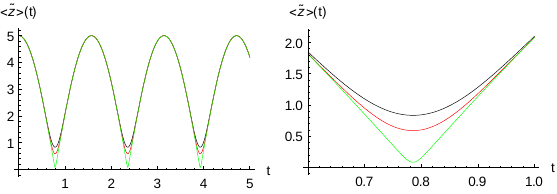}
\caption{
The scaled position expectation value \eqref{eq: zavHBBgaus} is plotted for a harmonic BB described initially 
by the Gaussian wavepacket \eqref{eq: psi0G} at different dynamical regimes: $ \ep = 1 $ (black curves), 
$ \ep = 0.5 $ (red curves) and $ \ep = 0.01 $ (green curves) for $\si_0 = 1$ and $ z_0 = 5 $; lengths are 
in units of  $ \ell_s = \sqrt{ \hb/ (2m\om) } $ and time in units of $t_s=2/\om$.}
\label{fig: zavHBBgaus}
\end{figure}
%
%===============================================

In Fig. \ref{fig: zavHBBgaus}, the  scaled position expectation value \eqref{eq: zavHBBgaus} is plotted 
for a harmonic BB described initially by the Gaussian wave packet \eqref{eq: psi0G} at different 
dynamical regimes. This quantity has a periodic structure. Since the initial wave function is the same for 
all dynamical  regimes then the initial value is also the same. As the right panel shows, when approaching 
the classical limit, bouncing from the hard wall occurs, on average, at a closer distance to the wall showing that
the quantum-classical transition is smooth and continuous. In the strict classical regime, the ball collides with the
wall.

\section{Ehrenfest relations and non-classical force}

In this section, we consider Ehrenfest relations and define the concept of non-classical force reminiscent 
of the quantum effective force which has already been introduced in the realm of quantum 
mechanics \cite{DoAn-PLA-2000, DoAn-LP-2002} and has been applied to time-dependent 
traps \cite{Mo-PS-2012, Mo-PS-2014}. For the variation of the position and momentum expectation 
values, one has that
\begin{eqnarray}
\frac{d}{dt} \wti{\la \hat{z} \ra}(t) &=& \frac{d}{dt} \int_0^{\infty} z | \tPsi(z, t) |^2 
= \frac{ \wti{\la \hat{p} \ra}(t) }{ m } , \label{eq: ehr1}
 \\
\frac{d}{dt} \wti{\la \hat{p} \ra}(t) &=& \frac{d}{dt} \int_0^{\infty} \tPsi^*(z, t) \frac{\thb}{i} \frac{\pa}{\pa z} \tPsi(z, t) \nonumber \\
&=& \bigg\la \tPsi(t)  \bigg| - \frac{\pa V(z)}{\pa z}  \bigg| \tPsi(t) \bigg\ra + \frac{\thb^2}{2m} \bigg| \frac{\pa}{\pa z} \tPsi(z=0, t) \bigg|^2   \label{eq: ehr2},
\end{eqnarray}
where we have used the standard boundary conditions.
%on the scaled wavefunction and its derivative and the STDSE. 
Note that the scaled wavefunction must be square integrable. Furthermore, its derivative must be zero at infinity. For the simple harmonic oscillator, the first term of Eq. \eqref{eq: ehr2} is proportional to the expectation value $ \wti{\la z \ra}(t) $.
The second extra term shows the effect of the hard wall at $ z=0 $. We call this extra term, in comparison to the customary Ehrenfest relation, the {\it non-classical} force since it contains non-classical effects only,
\begin{eqnarray} \label{eq: nc-force}
\ti{f}_{\nc}(t) &=& \ep \frac{\hb^2}{2m} \bigg| \frac{\pa}{\pa z} \tPsi(z=0, t) \bigg|^2   .
\end{eqnarray}
This effective force is always positive, a repulsive force, causing that the ``center of mass" of the 
nonclassical wave packet  moves far from the floor, in comparison to the case of without wall \cite{DoAn-PLA-2000}.
Classically, BB is affected by the wall only when it collides with the wall i.e, locally. But, this is not the case for other regimes. Note that $ \tPsi(z=0, t) $ itself depends on the transition parameter $\ep$. 
This effective force affects too the variance in momentum, 
$ \wti{\del p}(t) = \wti{\la \hat{p}^2 \ra} - \wti{\la \hat{p} \ra}^2 $,  in the following way 
\begin{eqnarray}
\frac{d}{dt} \wti{\del p}(t) &=& \im \left\{ 2 \thb ~ \int_0^{\infty} dz \frac{\pa V(z)}{\pa z} \tPsi(z, t) \frac{\pa}{\pa z} \tPsi^*(z, t)  
- \thb^3 ~ \frac{\pa \tPsi(z=0, t)}{\pa z} ~ \frac{\pa^2 \tPsi^*(z=0, t)}{\pa z^2} \right\} \nonumber \\
&-& 2 \wti{\la \hat{p} \ra}(t) ~ \frac{d}{dt} \wti{\la \hat{p} \ra}(t)   ,
\end{eqnarray}
where ``Im" stands for the ``imaginary part". Note that $\dot{\wti{\del p}}(t)$ is affected by the hard wall 
through two terms (i) $ - \thb^3 (\pa_z \tPsi) (\pa_z^2 \tPsi^*) |_{z=0} $ (ii) $ - 2 \wti{\la \hat{p} \ra}(t) \ti{f}_{\nc}(t) $.
One then obtains
\begin{eqnarray} \label{eq: nc-force-gauss}
\ti{f}_{\nc}(t) &=& \ep\frac{\hb^2}{2m}~ \frac{4 \sqrt{2} z_0^2 \si_0 m^3 \om^3   }{ [\ep \hb^2 \sin^2( \om t) + 4 m^2 \si_0^4 \om^2 \cos^2(\om t)]^{3/2} }
\exp \left[ - \frac{ 2 m^2 z_0^2 \si_0^2 \om^2 }{ 4 m^2 \si_0^4 \om^2 + \ep \hb^2 \tan^2( \om t) } \right]  ,
\end{eqnarray}
for the non-classical force when the initial packet is Gaussian, Eq.  \eqref{eq: psi0G}. This relation shows that $ \ti{f}_{\nc}(t) $ is a periodic function of time with period $ \pi / \om $.

\section{Conclusions}

The scaled quantum theory is used in this work to analyze the well-known  problem of a bouncing ball in 
a gravitational field and  harmonic potential. The theory has been presented in a very general way by 
considering first the Liouville-von Neumann equation for mixed states and afterwards it has been particularized 
to the case of pure states. The main purpose of this work is to show how this theory works for 
conservative systems and how the quantum-classical transition is carried out in a continuous and smooth
way; in-between dynamical regimes being considered as intermediate steps in this limiting process. 
%Interestingly enough, 
These in-between dynamical regimes still show quantum features which fade gradually to disappear completely when the quantumness parameter is strictly zero. 
%It is also shown that the scaled theory gives the right answers for any value of the quantum number considered whereas the WKB approximation is only good at higher and higher quantum numbers. Moreover, 
One also sees clearly how the locality of the 
wave function is increasing as we are approaching the classical limit and the difference between 
eigenvalues is also going to zero. The uncertainty and Ehrenfest relations are also discussed in terms 
of the scaled theory showing the right behavior again in the classical limit.

The scaled theory has been extensively applied to open quantum systems by ourselves in the past and with 
this new study we do think that the corresponding theory, in spite of the fact that gives a first order classical theory instead of the usual second order one, is in a situation to answer all the problems 
concerning the quantum-classical transition in terms of wave functions and density matrix. In our opinion,
this approach provides an alternative and complementary view of quantum and classical motions as well as 
in-between dynamical regimes. New studies in non-equilibrium statistical mechanics within this framework such as, for example,  the scaled Fokker-Planck 
or, in general, scaled diffusion  are interesting to follow.

\vspace{0.5cm}

{\bf Acknowledgements}:
SVM acknowledges support from the University of Qom and SMA support from the Fundaci\'on Humanismo 
y Ciencia.

\vspace{0.5cm}

{\bf Data Availability Statement}: This manuscript has no associated data.

%********************************************************

%===========================

%
\end{document}